# Investigating the Near-Infrared Properties of Planetary Nebulae. I. Narrowband Images

William B. Latter [1]

National Radio Astronomy Observatory, [2]

949 N. Cherry Ave. Campus Bldg. 65, Tucson, AZ 85721

Douglas M. Kelly

Department of Astronomy, University of Texas, Austin, TX 78712

Joseph L. Hora

Institute for Astronomy, 2680 Woodlawn Drive, Honolulu, HI 96822

Lynne K. Deutsch

Five College Astronomy Department, Univ. of Massachusetts, Amherst, MA 01003

---

[1] Current address: NASA/Ames Research Center, MS 245–3, Moffett Field, CA 94035

## ABSTRACT

We present the results of a near-infrared narrowband imaging survey of planetary nebulae. Objects were selected in a way that complements similar surveys done at visible and near-infrared wavelengths. No new detections of molecular hydrogen emission were made. The $H_2$ is frequently found to be extended, except in young, visibly compact objects. Our results are consistent with the already determined correlation of $H_2$ emission with planetary nebula morphological type. Filamentary and other kinds of structures are clearly resolved in many nebulae.

*Subject headings:* ISM: molecules – planetary pebulae: general – stars: AGB and post-AGB – infrared: ISM: continuum – ISM: structure

## 1. Introduction

Ever since their discovery more than 200 years ago, planetary nebulae (PNe) have been the subject of intense study. These beautiful and enigmatic regions have revealed many of their secrets, most within the past 20 years. Recent reviews of PNe and PNe evolution include Kwok (1994), Peimbert (1992), and Pottasch (1992). With the advent of wide-field array detectors operating in the near-infrared (near-IR) spectral region, a new and potentially critical tool for the study of PNe is now available. While PNe emit throughout the electromagnetic spectrum, the near-IR has many advantages over other wavelengths. In particular, it is well-known that PNe contain a substantial quantity of dust that was probably formed while the star was on the asymptotic giant branch (AGB). The near-IR therefore provides lower attenuation to important dense regions than the visible. Atmospheric seeing at near-IR wavelengths is typically reduced compared to that at visible wavelengths. When coupled to high-spatial resolution cameras, this can help to resolve structure that is not normally seen at the shorter wavelengths. The molecular distribution in PNe is not well understood, and many transitions from vibrationally excited $H_2$ and other molecular species fall in the near-IR. While the $H_2$ transitions are excited by UV photons in the photodissociation front of some younger PNe (see Hora & Latter 1994; Dinerstein et al. 1988), for the most part $H_2$ in PNe is excited by shocks (Kastner et al. 1994). Several studies have already been made that explore the near-IR properties of PNe, both as a group and of specific examples (e.g., Kelly & Latter 1995; Hora & Latter 1994; Kastner et al. 1994; Webster et al. 1988; Zuckerman & Gatley 1988).

It is for these reasons that we have undertaken the study to be discussed in this and following papers – a near-IR imaging and spectral study of a moderately sized sample of PNe. Our goals for this study were manyfold. Specifically, we have looked for systematic differences between the near-IR appearance of PNe to that in the visible (e.g., Schwarz et al. 1992; Balick 1987), as well as acquired new, high-spatial resolution data on extended $H_2$ emission. Because the near-IR transitions of $H_2$ are excited only in hot gas, or that exposed to intense UV fields, it probes the gas closest to the photodissociation region or that impacted by shocks. A focus of the study was on narrowband line imaging of $H_2$ and hydrogen Br$\gamma$. Our object selection was made so that there is considerable intentional overlap with previous CCD and $H_2$ surveys. It has been suggested that there is a strong correlation of molecular detection (particularly $H_2$) with PNe morphological type, and possibly progenitor mass (Huggins & Healy 1989; Zuckerman & Gatley 1988). To more fully understand this correlation, attempts were made to make new detections of extended $H_2$ emission as well as image objects previously detected. A full interpretation of these data combined with the results of a near-IR spectral survey currently underway will be presented in additional papers.

## 2. Observations

The data presented in this paper include images acquired from Kitt Peak, Arizona and Mauna Kea, Hawaii between 1992 June and 1993 December. The Kitt Peak observations were made with the Steward Observatory (SO) 2.3m telescope using the SO 256×256 HgCdTe NICMOS3 array camera and $\sim 1\%$ bandwidth tunable filters. The tilting narrowband filters were tuned to hydrogen Br$\gamma$ ($\lambda = 2.1655$ $\mu$m), $H_2$ $v = 1 \rightarrow 0$ S(1) ($\lambda = 2.1213$ $\mu$m), and adjacent continuum (see Latter et al. 1993 for details). Cold (77 K) reimaging optics provided an image scale of $0\farcs6$ pixel$^{-1}$. For some objects, broadband $J$, $H$, and $K$ images were also obtained. Weather conditions during much of the observing period were highly irregular, making precise flux calibration difficult. Since determining the near-IR morphologies of PNe was a main goal, we have not flux calibrated the suspect data. In a typical 15 minute integration (on + off) in the narrowband filters, we found a 3$\sigma$ sensitivity of $\approx 4.0 \times 10^{-16}$ ergs s$^{-1}$ cm$^{-2}$ pixel$^{-1}$ in $0\farcs6$ pixels. Most integration times for images presented here were 6 – 10 minutes on source. For some objects such as J 900, NGC 2346, and M 57, we integrated longer to bring out faint emission. The seeing at $\lambda = 2.2$ $\mu$m was in the range $0\farcs7 - 1\farcs5$, with $\sim 1\farcs1$ being typical.

Observations on Mauna Kea were made using the University of Hawaii (UH) 2.2m telescope and UH 256×256 HgCdTe NICMOS3 array camera (Hodapp et al. 1992) at an image scale of $0\farcs12$ pixel$^{-1}$ or $0\farcs24$ pixel$^{-1}$. Measurements were made in the $J$, $H$, and $K$ bandpasses, and fixed wavelength narrowband filters were used for observations of Br$\gamma$, H$_2$ $v = 1 \rightarrow 0$ S(1), and $\lambda = 2.23 - 2.29$ $\mu$m continuum. Because of the much broader bandpass of the UH continuum filter relative to the line filters, and possible H$_2$ line contamination, continuum subtraction was not attempted with the UH data. Observations were also made in the $I$–band using the UH 2048×2048 CCD camera. Seeing on Mauna Kea at $\lambda = 2.2$ $\mu$m was typically $0\rlap.{''}5 - 0\rlap.{''}7$. Therefore, these data provide much higher spatial resolution than the Kitt Peak data (on average). For this reason, we have included data from both sites for some of the same objects.

The images were constructed from many short exposures that were individually sky-subtracted and flat-fielded before being shifted and averaged to form final images. For extended sources, the telescope was nodded to nearby regions of blank sky to obtain sky frames. Compact sources could be observed efficiently by rastering them on the array. In this way, one object frame acted as the sky frame for the next. Flat-field frames were produced by averaging blank sky frames and then subtracting a dark frame of equal integration time. All of the data were reduced using standard IRAF routines. When accurate flux calibration was warranted, it was determined by comparison with standard stars from Elias et al. (1982). It is important to note that there can be a contribution from He I $3^3P^\circ - 4^3S$ at $\lambda = 2.114$ $\mu$m in the H$_2$ narrowband images. In virtually all cases, such a contribution can be determined from the location of the emission. We will refer to this potential problem on a case-by-case basis. The data are summarized in Table 1, and filter parameters are listed in Table 2. The grayscale images are presented in Figures 1 – 15 (Plates # – #). Note that in the figure captions, UH data is specifically identified as such. Typically, four images are presented for each object selected from the following: $I$–band, $J$–band, $H$–band, $K$–band, H$_2$, Br$\gamma$, narrowband continuum, continuum subtracted H$_2$, and continuum subtracted Br$\gamma$.

## 3. Discussion

We will break up our discussion into overall results and confirmed H$_2$ emission objects. Note that whenever possible, the morphological types listed in Table 1 and mentioned below are those of Balick (1987), or they follow the convention defined by him. A difficulty in morphological classification is that it can depend on the wavelength observed and flux level considered. In this paper, we will use the term "bipolar reflection nebula" to refer only to those post-AGB objects and proto-PNe (PPNe) that display a double-lobed appearance at visible wavelengths and have the characteristic that the lobes are seen primarily in reflected light from the central star (such as AFGL 2688). A "butterfly" or "bipolar" nebula is a PN that is bi-lobed and has an equatorial waist (e.g., M 2–9; see Schwarz et al. 1993).

These types are likely the same general structure viewed at different evolutionary states (such that a bipolar reflection nebula evolves to butterfly), but this relationship has not yet been proved. In general, we will use the term "polar lobes" to refer to those regions that are identifiable as the "wings" of butterfly-type nebulae or the lobes of the younger bipolar reflection nebulae.

### 3.1. General Results

The data in this paper represent a good selection of PNe morphological types. While many interesting features and structures have been found in a number of these objects, no clear systematic differences between visible light images and the near-IR are seen. This is meant in the sense that nebular classification in the visible is, in most cases, the same as that found in the near-IR. Differences seen are generally related to excitation of various species and the location of shocks. The Br$\gamma$ emission follows H$\alpha$ closely in objects where differential attenuation is not a factor, with apparent exceptions discussed below. The continuum in our $\sim 1\%$ wide filters can be significant, but typically the near-IR emission from PNe is dominated by Br$\gamma$. For the most part, we have found that the near-IR continuum tends to mimic the Br$\gamma$ distribution, and continuum subtraction does not offer significantly different morphological information. Given our sensitivity limits, extended halos of the type seen by Balick et al. (1992) could not be detected in Br$\gamma$.

Detailed structure is seen in all bands imaged. Bright knots or clumps can be found in most objects in both line and continuum emission. Filamentary structures have been resolved in a number of the PNe. Some PNe have relatively smooth halos surrounding complex inner structures. A transition from smooth, uniform mass loss to a much more dynamic process is indicated. The effects generated by the interaction of multiple velocity winds is clearly evident in some objects (e.g., NGC 7009). Others have structures that are difficult to unravel into a three-dimensional picture (e.g., NGC 6543, NGC 6826). In the following, we discuss several sources that were either not observed in H$_2$ or in which there is no clear evidence for H$_2$ emission.

#### 3.1.1. NGC 6302

We have displayed images only of the central region of NGC 6302 (Fig. 5b), which is extended in the visible beyond the array limits (see Schwarz et al. 1992). However, data taken toward regions outside that shown indicate little extended emission, likely owing to our sensitivity limits. Emission seen in the H$_2$ and the Br$\gamma$ filters are very similar, while the continuum appears much less extended. This object has a very hot central star, and as such, it is unlikely that the emission observed in the 2.12 $\mu$m filter is from H$_2$, but is more likely He I $3^3$P$^\circ - 4^3$S contamination. The $J$–band image appears to have the same characteristics as the 2.26 $\mu$m continuum. However, a pronounced "arc" of emission is present on the south–east side, which does not appear so strongly in the other data.

Lester & Dinerstein (1984) have presented maps of NGC 6302 in the $J-$ and $K-$bands, together with longer wavelengths. They reported the detection of an infrared disk at the center of the nebula that might have been associated with the mechanism for collimating the outflow of material into the polar lobes. However, it is apparent from the higher resolution images presented here that the distribution of near-infrared emission is similar to the visible appearance, at least in the central region. In particular, the dark lane that is found in visible images is also seen in the $J-$ and $K-$bands. Therefore, we find no evidence for a disk or toroidal structure associated with the mechanism for shaping the nebula.

### 3.1.2. NGC 6572

NGC 6572 has been imaged previously in the visible (e.g., Jewitt et al. 1986; Schwarz et al. 1992). However, our near-IR images reveal much more detail (Fig. 6b). This object appears to be of the bipolar or early butterfly morphological type but is not compressed at the equator, so it is better to classify it an elliptical. There is an intensity enhancement in the equatorial region, however. Johnson & Jones (1991) found NGC 6572 to be polarized in the visible with $P \sim 1.4\%$. This would suggest that at least part of the light observed in the near-IR is polarized as well. Indeed, there are indications of horn–like structures of the type seen in AFGL 2688 (see Fig. 14a). These features in AFGL 2688 have been modeled and found to be a result of single-photon scattering in an axisymmetric distribution of dust (Latter et al. 1993; see also Morris 1981). There is some evidence in the H$_2$ filter image for He I emission that appears as a bright spot to the northwest of the central star and is not seen in the other data.

### 3.1.3. NGC 7026

Emission at the location of the bright knots in our H$_2$ continuum subtracted images (Fig. 8b) is likely from He I. NGC 7026 appears similar in general morphology to NGC 6572. Balick (1987) classified this object as a late elliptical.

### 3.1.4. BD+30°3639

This young, carbon-rich PN is similar in many respects to NGC 7027. Both are elliptical (in the infrared; Fig. 10) with strong near- and mid-IR excesses. Both have $H_2$ emission in the halo regions, and the so-called unidentified infrared (UIR) emission features are prominent at 3.28 $\mu$m and in the mid-infrared (we did not image BD+30°3639 in $H_2$). BD+30°3639 has been imaged previously in the near-IR (Graham et al. 1993a; Hora et al. 1993; Roche 1989; Smith et al. 1989). The $K$-band image presented here is at higher spatial resolution and shows more details of the nebular structure. The nebula appears slightly rectangular, with the brightest portion of the lobes in the N and S regions. The central star is well separated from the ring at this resolution, showing clearly the evacuated central region. The upper right frame in Fig. 10b reveals a faint halo that extends out to about twice the diameter of the bright ring. The $H_2$ emission detected by Graham et al. (1993a) is visible at the E edge of the halo, as well as several other regions along the edge of the halo.

### 3.1.5. IC 418

IC 418 is a young, low-excitation PNe (Fig. 11b) and has previously been imaged in the near-IR (Hora et al. 1993). Like BD+30°3639 and NGC 7027, it exhibits the UIR features in the near- and mid-IR. The near-IR spectrum is dominated by lines of H and He, with a hot dust continuum (Hodapp et al. 1994; Zhang & Kwok 1992). In general, these new data confirm the previous imaging results. In particular, the $J$-band image is of higher resolution and greater sensitivity than previous images. The halo that is detected in the $H$- and $K$-band images is also detected at $J$, and extends $\sim 30''$ in diameter. One discrepancy between this image and the $J$-band image in Hora et al. is that the previous $J$ image showed the east lobe significantly rotated towards the north. The location of the east lobe in the new $J$ image is similar to the lobe position in the $H$- and $K$-bands, and in Br$\gamma$ and H$\alpha$ images (Louise et al. 1987).

### 3.1.6. M 1–92

M 1–92 ("Minkowski's Footprint") is a proto-typical example of a bipolar reflection nebula. In the visible, the fully obscured B0.5 V (Herbig 1975) central star is seen only in reflected starlight scattered off two symmetrical bipolar lobes (Schmidt et al. 1978). The near-IR appearance is significantly different from that seen in the visible, however (Fig.

14b). Our near-IR images show that M 1–92 bears a striking resemblance to AFGL 618, which is also a bipolar proto-planetary nebula (see Fig. 13b and Latter et al. 1992). As in that object, M 1–92 has one bright peak situated roughly between the two lobes. The western lobe is brighter than the east, and the difference is much more pronounced at $J$ than at $K$. The bright peak is not seen at visible wavelengths (Schmidt et al. 1978). The morphology displayed is consistent with the single-photon scattering models of Morris (1981) and Latter et al. (1992) as applied to AFGL 618. This implies that the polar axis of the nebula is highly inclined to the plane of the sky ($i \sim 45°$) with the western lobe being closest to the earth.

## 3.2. Detected $H_2$ Emission Objects

We imaged several objects not observed in the $H_2$ surveys of Zuckerman & Gatley (1988) and Webster et al. (1988), but no new $H_2$ detection was clearly made. We will, however, be better able to assess the $H_2$ content of PNe when we discuss the spectroscopic results in a following paper. The images of M 2–9 and AFGL 2688 acquired during the course of this survey have already been discussed at length elsewhere (Hora & Latter 1994; Latter et al. 1993) and will not be considered further here.

### 3.2.1. NGC 2346

First mapped in $H_2$ by Zuckerman & Gatley (1988) and recently imaged by Kastner et al. (1994), NGC 2346 has a remarkable butterfly–type morphology surrounding a bright elliptical region (Fig. 2b). The appearance of the lobes is like that of limb-brightened bubbles. It is likely that the $H_2$ emission arises from shocked gas at the interface of an expanding bubble of fast-moving, low-density gas interacting with the slower moving red giant envelope. We find the continuum to be weak, and it appears to follow the butterfly structure. Br$\gamma$ emission is very centrally concentrated, with no detectable emission from the more extended elliptical region. This is different from the H$\alpha$ distribution, which fills the central elliptical region (Balick 1987). This difference is likely, though not entirely, a result of insufficient sensitivity. A large amount of filamentary structure is visible in the $H_2$ emission. It has been argued, based largely on the $H_2$ distribution, that this object and M 57 (§3.2.2.) have very similar overall morphologies with M 57 being viewed nearly pole-on (Kastner et al. 1994; see also Balick et al. 1992). A contour plot of the $H_2$ emission is shown in Fig. 16. The appearance of NGC 2346 would suggest that the object is inclined by a few degrees such that the north lobe is pointed toward the earth.

### 3.2.2. M 57

M 57 (NGC 6720; the Ring Nebula) has been studied in molecular hydrogen emission by several groups (Kastner et al. 1994; Greenhouse et al. 1988; Zuckerman & Gatley 1988). The main ring of this object is a strong source of $H_2$ emission, and has modest Br$\gamma$ and continuum. Independent of Kastner et al. (1994), we found the emitting $H_2$ to extend well beyond the bright ring. This emission is coextensive with the visible halo, but detailed inspection finds that the $H_2$ emission is slightly offset from H$\alpha$ (see Kastner et al. 1994). The Br$\gamma$ emission primarily falls interior to the $H_2$, while the continuum follows the $H_2$ emission. A large amount of detailed filamentary structure is visible in our grayscale $H_2$ image (Fig. 13a). These filaments point radially away from the main ring, and in at least a few cases seem to originate from bright knots in the ring. Such filaments might, therefore, reflect higher density regions created by instabilities in the flow. These data are consistent with the idea that M 57 is a butterfly-type nebula seen pole-on. The extent of the $H_2$ emission is clearly visible in Fig. 17. Kastner et al. (1994) argue convincingly that the $H_2$ is shock excited, but we will consider this further in our forthcoming paper. A detailed study of the $H_2$ velocity structure in this object and NGC 2346 could be very illuminating.

### 3.2.3. NGC 7027

This young compact PN has a bright continuum that is largely coincident with a ring of Br$\gamma$ emission. The near-IR morphology of NGC 7027 has been studied previously (Kastner et al. 1994; Graham et al. 1993b). Our $H_2$ image (Fig. 10a) has a strong contribution from He I emission, which lies in a ring inside that of the $H_2$. Graham et al. (1993a) argue on morphological grounds that the $H_2$ emission must be primarily excited by UV photons, rather than being shock excited. Contour diagrams of continuum subtracted $H_2$ and Br$\gamma$ images are shown in Fig. 18. The calibration of these images agrees extremely well with that of Graham et al. (1993b), lending confidence to our continuum subtraction technique.

### 3.2.4. J 900

Our near-IR Br$\gamma$ continuum subtracted image of J 900 (Figs. 12b and 19) is elongated into two lobes but otherwise featureless, much like the morphology seen in H$\alpha$ and [O III] (see Schwarz et al. 1992). The continuum is elongated in the direction of the lobes. In contrast, we find $H_2$ emission to extend from the central region in almost jet-like structures.

At the end of the the north-west "jet" appears an arc of emission. It is likely that He I emission contributes to our H$_2$ continuum subtracted image, with a spatial distribution similar to that of Br$\gamma$. This object has been discussed recently by Shupe et al. (1995).

### 3.2.5. NGC 2440

NGC 2440 has extended H$_2$ emission that does not obviously follow the structure seen in visible CCD images (Cutri 1994; Schwarz et al. 1992; Balick 1987). We find that the H$_2$ emission extends in roughly linear structures from the center, in between the visible lobes, and ends in arcs that appear to almost encircle the nebula (Figs. 4a and 20). The arcs are not equidistant from the central region, an effect that is likely caused by orientation to the line of sight. Bright knots visible in our H$_2$ continuum subtracted image, which are approximately coincident with the Br$\gamma$ emission, are from He I emission in the filter bandpass.

### 3.2.6. AFGL 618

AFGL 618 (Fig. 13b) is dominated by continuum emission in the visible and near-IR; arising primarily from scattered light emitted by the central star. It is a source of strong H$_2$ emission, but the distribution of that emission has been unknown because of the strong continuum (see, however, Rowlands 1993). We have successfully subtracted the continuum from our H$_2$ and Br$\gamma$ images. The emitting H$_2$ is concentrated in the lobes (Fig. 21), similar to that found for AFGL 2688. No equatorial H$_2$ emission like that seen in AFGL 2688 is detected, however. How the H$_2$ might be excited in AFGL 618 and AFGL 2688 has been discussed by Hora & Latter (1994), Latter et al. (1993; 1992), and Jura & Kroto (1990).

Br$\gamma$ is confined to an unresolved, centrally located region (Fig. 21). The position and size of the ionized region has been discussed by Kwok & Bignell (1984). They find it to be $\sim 0\rlap{.}''4$ in diameter, and located roughly between the two visible lobes. This is consistent with our Br$\gamma$ data. The integrated fluxes are $f(\mathrm{H}_2) = 1.1 \pm 0.33 \times 10^{-12}$ ergs s$^{-1}$ cm$^{-2}$ and $f(\mathrm{Br}\gamma) = 1.5 \pm 0.45 \times 10^{-13}$ ergs s$^{-1}$ cm$^{-2}$. These values agree very well with spectroscopic results (Latter et al. 1992; Kelly, Latter, & Rieke 1992; Thronson 1981).

Broadband integrated magnitudes for these data are $J = 13.7 \pm 0.2$ mag, $H = 11.4 \pm 0.2$ mag, and $K = 8.9 \pm 0.2$ mag. Conditions during the broadband measurements were not photometric, but we consider these data to be in agreement with that of Latter et al.

(1992), who found a significant discrepancy with previous $K$–band measurements. Our new $J$–band image is consistent with the scattering model from that study.

### 3.3. Molecular Hydrogen Distribution in Planetary Nebulae

Emission from low-lying $H_2$ electric quadrupole vibration–rotation transitions can be excited collisionally in regions with kinetic temperatures $T_K \gtrsim 1000$ K. Kinetic temperatures like this are typically thought to be associated with the presence of moderate velocity shocks. For shocks to produce observable $H_2$ emission the velocity must fall in the range $V_s \approx 5 - 25$ km s$^{-1}$, or somewhat higher for $C$–type shocks ($V_s \approx 40 - 50$ km s$^{-1}$; see Draine & Roberge 1982; Hollenbach & Shull 1977). Such velocities are observed in PNe, and $H_2$ excited in this way must be present at least part of the time. Ultraviolet photons with $\lambda > 912$ Å can escape from the ionized part of the nebula and then be absorbed in the Lyman and Werner bands of any $H_2$ in that region. About 10% of these absorptions will result in dissociation of the $H_2$ molecules. The remainder will lead to a cascade in the electronic ground state vibration-rotation levels, producing a near-IR spectrum. While this type of excitation can result in an easily identifiable spectrum (Black & van Dishoeck 1987), high densities and strong UV fields can produce a spectrum that mimics that of shock or thermal excitation (Sternberg & Dalgarno 1989). Often without detailed spectral information (Hora & Latter 1994; Ramsay et al. 1993), it can be difficult to know which is the principal cause of the excitation. In any case, it is instructive to isolate the regions where $H_2$ is emitting in PNe.

When present, it is evident that the emitting $H_2$ is always associated with the polar lobes of the nebulae, although it can also be present in other regions. In the case of bipolar reflection nebulae, such as AFGL 618 and AFGL 2688, the $H_2$ emission is concentrated within the visible lobes themselves (the equatorial emission in AFGL 2688 is discussed by Latter et al. 1993). In evolved butterfly-type nebulae (here we consider M 57 to be a member of this class), the $H_2$ emission comes from the outside edge of the lobes (see also, Kastner et al. 1994). It would appear that a transition occurs from emission within the lobe interior to the lobe surface, as the polar regions expand and evolve. This is the type of evolution that might be expected for wind-blown bubbles.

For other objects discussed here, it is not clear how the $H_2$ emission is associated with the overall distribution of matter. While not spherically symmetric in its $H_2$ emission, NGC 7027 does not obviously fall into the same class as the young bipolar reflection nebulae or the evolved butterfly nebulae. Its status as a very young, rapidly evolving object makes it likely that it should not be included in any specific evolutionary scenario of particular

– 12 –morphologies. A similar statement can be made about BD+30°3639, which has been detected in $H_2$ elsewhere (Graham et al. 1993a; Webster et al. 1988; Zuckerman & Gatley 1988). The cases of J 900 and NGC 2440 are even more difficult to interpret. Deeper $H_2$ images are required. Visible light images of NGC 2440 show it to be a butterfly nebula with lobes extending in the east–west direction (Cutri 1994; Schwarz et al. 1992). The $H_2$ emission is brightest between the lobes, in apparent contradiction to the trend discussed in the previous paragraph. In J 900 the $H_2$ is detected well beyond the visible manifestation of the nebula. The $H_2$ in that object extends outward in the direction to which the bright lobes are oriented (see Shupe et al. 1995).

A number of studies have pointed out a strong correlation of detectable molecular emission with PN morphological type (e.g., this work; Kastner 1994; Huggins & Healy 1989; Zuckerman & Gatley 1988). From these studies, it has been made clear that butterfly or bipolar morphological types appear to have much larger quantities of molecular material. Our new data are consistent with this result – if a PN is detected in $H_2$, it is a young bipolar reflection nebula or a more evolved butterfly nebula. As mentioned above, the exceptions consist of intermediate age objects that are rapidly dissociating the remnant AGB envelope. Note that the converse is not necessarily true – butterfly nebulae do not *always* have detectable $H_2$ or CO. This, however, might be more a statement of degree rather than contradiction. As emphasized by Huggins & Healy (1989), this correlation is very significant. Since the origin of nebular shapes, in particular bipolar and butterfly shapes, is the subject of much debate, it seems that this correlation must be investigated in some detail. The implications are of key importance. For example, for PNe to retain molecular envelopes at times significantly beyond exposure of the hot core, a minimum requirement is that large amounts of material must have been deposited into the molecular envelope. This would imply a strong correlation of mass loss rate on the AGB with morphological type. Also, the above mentioned molecular studies pointed out a clear trend for butterfly morphological types to be members of a young disk population, suggesting massive progenitor stars. Stars at the higher end of the AGB mass limit are capable of losing mass at high rates for an extended period of time before exposure of the core. It therefore appears that massive AGB stars produce butterfly nebulae. Thus, the close binary hypothesis for causing mass loss to be preferentially in the orbital plane might not be a necessary requirement (Zuckerman & Gatley 1988) unless all of these type of morphologies arise from binary systems with massive components (Han et al. 1995).

## 4. Summary

We have presented the results of a near-IR imaging survey of a moderately sized sample of planetary nebulae. Aside from the occasional presence of $H_2$ emission, there are no systematic morphological differences with PNe observed in the visible, the exceptions being very dusty objects like AFGL 618 and AFGL 2688. An attempt was made to make new detections of molecular hydrogen emission, without success. However, new images of several $H_2$ emission objects are presented. With the exception of young, compact objects, the $H_2$ emission is typically found to be extended and does not always trace visible components of the nebulae. There is, however, some ambiguity when interpreting compact emission that might be confused with He I emission in the filter bandpass. Our new high-spatial resolution data reveal filamentary and other kinds of structures that are clearly resolved in many nebulae. Previous studies (e.g., Zuckerman & Gatley 1988) determined a correlation of $H_2$ emission with "butterfly" morphological type. Our results are consistent with this finding. It has been suggested that molecular material is only present in objects with massive progenitors (Huggins & Healy 1989; Zuckerman & Gatley 1988). This might well be the case, but the strong correlation with the morphology of PNe is still not explained, unless morphology is also related to progenitor mass and mass-loss rate (see Corradi & Schwarz 1994). The data presented in this paper will be further analyzed and combined with near-IR spectroscopy in a following paper.

The authors are indebted to George Rieke, who provided numerous suggestions and assistance during some of the observations. Valuable comments from the referee, Hugo Schwarz, and Jim Liebert improved the presentation of this work. We are grateful to Richard Wainscoat for use of the UH CCD camera to obtain the $I$-band images. We thank John Bieging and Phil Jewell for interesting and motivating discussions. Marcia Rieke and Roc Cutri are kindly thanked for helpful conversations. We are especially gratefully to Dennis Means for his patience and ability to keep us awake during the many rainy nights seen by this project. We also thank the SO TAC for generous and repeated allocations of time for a project that came to be known as "The Rainmaker." W.B.L. thanks NRAO for support through a Jansky Fellowship. D.M.K. acknowledges support from NSF grants AST90–20292 and AST91–16442. We also thank E. F. Montgomery for his help with setting up the narrowband filter system in the SO camera.

Table 1
Summary of Observations

| # | Object | PN G #[1] | Type[2] | SO Filters[3] | SO Obs. dates | UH Filters[3] | UH Obs. dates | Fig. # |
|---|--------|-----------|---------|---------------|---------------|---------------|---------------|--------|
| 1 | NGC 40 | 120.0+09.8 | mid. elliptical | K, $H_2$, $Br\gamma$, cont | 15/10/92 | — | — | 1a |
| 2 | NGC 1535 | 206.4−40.5 | early round | K, $H_2$, $Br\gamma$, cont | 15/10/92, 29/12/93 | — | — | 1b |
| 3 | NGC 2022 | 196.6−10.9 | early elliptical | $H_2$, $Br\gamma$, cont | 28/10/93, 29/12/93 | — | — | 2a |
| 4 | NGC 2346 | 215.6+03.6 | mid. butterfly | K, $H_2$, $Br\gamma$, cont | 14&15/10/92, 10/3, 27/10 93 | — | — | 2b |
| 5 | NGC 2371-2 | 189.1+19.8 | late elliptical | $H_2$, $Br\gamma$, cont | 29/12/93 | — | — | 3a |
| 6 | NGC 2392 | 197.8+17.3 | early round | Ks, $H_2$, $Br\gamma$, cont | 7&9/3/93 | — | — | 3b |
| 7 | NGC 2440 | 234.8+02.4 | late butterfly | Ks, $H_2$, $Br\gamma$, cont | 10/3/93 | — | — | 4a |
| 8 | NGC 3242 | 261.0+32.0 | early elliptical | Ks, $H_2$, $Br\gamma$, cont | 10/3/93, 29/12/93 | — | — | 4b |
| 9 | NGC 6210 | 043.1+37.7 | irregular | Ks, $H_2$, $Br\gamma$, cont | 10/3/93 | — | — | 5a |
| 10 | NGC 6302 | 349.5+01.0 | mid. butterfly[4] | — | — | J, K, $H_2$, $Br\gamma$, cont | 21&22/6/92, 28/6/93 | 5b |
| 11 | NGC 6543 | 096.4+29.9 | peculiar+round halo | — | — | $H_2$, $Br\gamma$, cont, I | 19/6/92, 28/6/93 | 6a |
| 12 | NGC 6572 | 034.6+11.8 | late elliptical[4]? | — | — | K, $H_2$, $Br\gamma$, cont | 20/6/92, 28/6/93 | 6b |
| 13 | NGC 6826 | 083.5+12.7 | early elliptical +round halo | — | — | I | 19/6/92 | 8a |
| 14 | NGC 7009 | 037.7−34.5 | mid. elliptical | K, $H_2$, cont | 15/10/92 | $H_2$, $Br\gamma$, cont | 27&28/6/93 | 7 |
| 15 | NGC 7026 | 089.0+00.3 | late elliptical | K, $H_2$, $Br\gamma$, cont | 15/10/92, 27/10/93 | I | 19/6/92 | 8 |
| 16 | NGC 7027 | 084.9−03.4 | mid. elliptical | K, $H_2$, $Br\gamma$, cont | 15/10/92, 27/10/93 | — | — | 10a |
| 17 | NGC 7662 | 106.5−17.6 | early elliptical | K, $H_2$, $Br\gamma$, cont | 14/10/92, 27/10/93 | I, $Br\gamma$, cont | 28/6/93 | 9 |

Table 1 *(continued)*
Summary of Observations

| # | Object | PN G #[1] | Type[2] | SO Filters[3] | SO Obs. dates | UH Filters[3] | UH Obs. dates | Fig. # |
|---|---|---|---|---|---|---|---|---|
| 18 | BD+30°3639 | 064.7+05.0 | early round | — | — | K | 20/6/92 | 10b |
| 19 | I21282+5050 | — | peculiar[4] | — | — | K | 20/6/92, 10/10/92 | 10b |
| 20 | IC 289 | 138.8+02.8 | mid. elliptical | K, $H_2$, Br$\gamma$, cont | 14/10/92, 28/10/93 | — | — | 11a |
| 21 | IC 418 | 215.2−24.2 | early elliptical | H, K, $H_2$, Br$\gamma$, cont | 15/10/92, 10/3/93 | J | 8/2/93 | 11b |
| 22 | IC 2149 | 166.1+10.4 | peculiar | $H_2$, cont | 28/12/93 | — | — | 11a |
| 23 | IC 4593 | 025.3+40.8 | irregular | Ks, $H_2$, Br$\gamma$, cont | 9/3/93 | — | — | 12a |
| 24 | J 900 | 194.2+02.5 | early butterfly[4]? | K, $H_2$, Br$\gamma$, cont | 14/10/92, 28/10, 28&29/12/93 | — | — | 12b |
| 25 | M 57 | 063.1+13.9 | mid. butterfly[5] | K, $H_2$, Br$\gamma$, cont | many | — | — | 13a |
| 26 | AFGL 618 | 166.4−06.5 | bipolar[4] | J, H, Ks, $H_2$, Br$\gamma$, cont | 9/3/93 | — | — | 13b |
| 27 | AFGL 2688[6] | — | bipolar[4] | J, H, K, $H_2$, cont | many | I, H, K, $H_2$, cont | 19&22/6/92 | 14a |
| 28 | M 1−78 | 093.5+01.4 | mid. butterfly[4] | — | — | K | 21/6/92 | 15b |
| 29 | M 1−92 | — | bipolar[4] | — | — | J, K | 22/6/92 | 14b |
| 30 | M 2−9[7] | 101.8+18.0 | mid. butterfly | K, $H_2$, cont | 13/6/92, 9/3/93 | J, H, K, $H_2$, Br$\gamma$, cont | 20-22/6/92, 27/6/93 | 15a |

[1] Strasbourg–ESO Planetary Nebula catalog number (Acker et al. 1992).
[2] From Balick (1987) for objects which appear in both data sets (see text).
[3] Pixel scale for UH NICMOS3 data is 0″.12 for data before 1993 June 1 and 0″.24 after. SO data is all at 0″.6/pixel. *I*–band data is at 0″.08/pixel.
[4] Not in Balick (1987).
[5] Identified as a middle elliptical+round halo by Balick (1987).
[6] See Latter et al. (1993).
[7] See Hora & Latter (1994).

**Table 2**
Filter Parameters

| Instrument | Filter | Half-Power Wavelengths ($\mu$m) |
|---|---|---|
| SO NICMOS3 Camera: | J | $1.258 \pm 0.14$ |
| | H | $1.647 \pm 0.17$ |
| | K | $2.205 \pm 0.21$ |
| | Ks | $1.990 - 2.320$ |
| | $H_2$ | $2.121 \pm 0.01$ |
| | continuum | $2.097 \pm 0.01$ |
| | Br$\gamma$ | $2.166 \pm 0.01$ |
| | continuum | $2.151 \pm 0.01$ |
| UH NICMOS3 Camera: | J | $1.238 \pm 0.14$ |
| | H | $1.686 \pm 0.14$ |
| | K | $2.211 \pm 0.20$ |
| | $H_2$ | $2.1132 - 2.1367$ |
| | Br$\gamma$ | $2.1646 - 2.2053$ |
| | continuum | $2.230 - 2.290$ |
| UH CCD Camera: | I | $0.816 \pm 0.091$ |

# REFERENCES

Acker, A, Marcout, J., Ochsenbein, F., Scholn, C., Stenholm, B., & Tylenda, R. 1992, The Strasbourg–ESO Catalogue of Galactic Planetary Nebulae (München: ESO)

Balick, B. 1987, AJ, 94, 671

Balick, B., Gonzalez, G., Frank, A., & Jacoby, G. 1992, ApJ, 392, 582

Black, J. H., & van Dishoeck, E. F., 1987, ApJ, 322, 412

Corradi, R. L. M., & Schwarz, H. E. 1995, A&A, 293, 871

Cutri, R. M. 1994, private communication

Dinerstein, H. L., Lester, D. F., Carr, J. S., & Harvey, P. M. 1988, ApJ, 327, L27

Draine, B. T., & Roberge, W. G., 1982, ApJ, 259, L91

Elias, J. H., Frogel, J. A., Mathews, K., & Neugebauer, G. 1982, AJ, 87, 1029

Graham, J. R., Herbst, T. M., Matthews, K., Neugebauer, N., Soifer, T., Serabyn, E., & Beckwith, S. 1993a, ApJ, 408, L105

Graham, J. R., Serabyn, E., Herbst, T. M., Matthews, K., Neugebauer, G., Soifer, B. T., Wilson, T. D., & Beckwith, S. 1993b, AJ, 105, 250

Greenhouse, M. A., Haywood, T. L., & Thronson, H. A. 1988, ApJ, 325, 604

Herbig, G. H. 1975, ApJ, 200, 1

Han, Z., Podsiadlowski, P., & Eggleton, P. P. 1995, MNRAS, in press

Hodapp, K.-W., Hora, J. L., Irwin, E., & Young, T. 1994, PASP, 106, 87

Hodapp, K.-W., Rayner, J., & Erwin, E. 1992, PASP, 104, 441

Hollenbach, D. J., & Shull, J. M. 1977, ApJ, 216, 419

Hora, J. L., Deutsch, L. K., Hoffmann, W. F., Fazio, G. G., & Shivanandan, K. 1993, ApJ, 413, 304

Hora, J. L., & Latter, W. B. 1994, ApJ, 437, 281

Huggins, P. J., & Healy, A. P. 1989, ApJ, 346, 201

Jewitt, D. C., Danielson, G. E., & Kupferman, P. N. 1986, ApJ, 302, 727

Johnson, J. J., & Jones, T. J. 1991, AJ, 101, 1735

Jura, M., & Kroto, H. 1990, ApJ, 351, 222

Kastner, J. H., Gatley, I., Merrill, K. M., Probst, R., & Weintraub, D. A. 1994, ApJ, 421, 600

Kelly, D. M., & Latter, W. B. 1995, AJ, 109, 1320

Kelly, D. M., Latter, W. B., & Rieke, G. H. 1992, ApJ, 395, 174

Kwok, S. 1994, PASP, 106, 344

Kwok, S., & Bignell, R. C. 1984, ApJ, 276, 544

Latter, W. B., Hora, J. L., Kelly, D. M., Deutsch, L. K., & Maloney, P. R. 1993, AJ, 106, 260

Latter, W. B., Maloney, P. R., Kelly, D. M., Black, J. H., Rieke, G. H., & Rieke, M. J. 1992, ApJ, 389, 347

Lester, D. F., & Dinerstein, H. L. 1984, ApJ, 281, L67

Louise, R., Macron, A., Pascoli, G., & Maurice, E. 1987, A&AS, 70, 201

Morris, M. 1981, ApJ, 249, 572

Peimbert, M. 1992, in Observational Astrophysics, ed. R. E. White (New York: IOP), 1

Pottasch, S. R. 1992, A&A Rev., 4, 215

Ramsay, S. K., Chrysostomou, A., Geballe, T. R., Brand, P. W. J. L., & Mountain, M. 1993, MNRAS, 262, 695

Roche, P. F. 1989, in IAU Symposium 131, Planetary Nebulae, ed. S. Torres–Peimbert (Dordrecht: Kluwer), 117

Rowlands, N. 1993, in Astronomical Infrared Spectroscopy: Future Observational Directions, ed. S. Kwok (ASP Conference ser. 41), 127

Schmidt, G. D., Angel, J. R. P., & Beaver, E. A. 1978, ApJ, 219, 477

Schwarz, H. E., Corradi, R. L. M., & Melnick, J. 1992, A&AS, 96, 23

Schwarz, H. E., Corradi, R. L. M., & Stanghellini, L. 1993, in IAU Symposium 155, Planetary Nebulae, eds. R. Weinberger and A. Acker (Dordrecht: Reidel), 214

Shupe, D. L., Armus, L. Matthews, K., & Soifer, B. T. 1995, AJ, 109, 1173

Smith, M. G. Geballe, T. R., Aspin, C., McLean, I. S., & Roche, P. F. 1989, in IAU Symposium 131, Planetary Nebulae, ed. S. Torres–Peimbert (Dordrecht: Kluwer), 178

Sternberg, A., & Dalgarno, A. 1989, ApJ, 338, 197

Thronson, H. A. 1981, ApJ, 248, 984

Webster, B. L., Payne, P. W., Storey, J. W. V., & Dopita, M. A. 1988, MNRAS, 235, 533

Zhang, C. Y., & Kwok, S. 1992, ApJ, 385, 255

Zuckerman, B., & Gatley, I. 1988, ApJ, 324, 501

**Figure 1:** a) NGC 40 in $\lambda = 2.151$ $\mu$m continuum, $K$–band, Br$\gamma$, and continuum subtracted Br$\gamma$. H$_2$ is not detected. A 20″ scale bar is shown. For this, and all images that follow, north is up and east is to the left. b) As in a) but for NGC 1535.

**Figure 2:** a) NGC 2022 in H$_2$, $\lambda = 2.097$ $\mu$m continuum, Br$\gamma$, and continuum subtracted Br$\gamma$. H$_2$ is not detected. A 20″ scale bar is shown. b) As in a) but for NGC 2346. The H$_2$ and continuum images are displayed on a logarithmic intensity scale. H$_2$ is clearly detected in an extended butterfly-type morphology. A 40″ scale bar is shown.

**Figure 3:** a) NGC 2371-2 shown in H$_2$, continuum subtracted H$_2$, Br$\gamma$, and continuum subtracted Br$\gamma$. H$_2$ is not detected. A 20″ scale bar is shown. b) NGC 2392 in H$_2$ $\lambda = 2.151$ $\mu$m continuum, Br$\gamma$, and continuum subtracted Br$\gamma$. H$_2$ is not detected. A 20″ scale bar is shown.

**Figure 4:** a) NGC 2440 shown in H$_2$, continuum subtracted H$_2$, $\lambda = 2.151$ $\mu$m continuum, and continuum subtracted Br$\gamma$. H$_2$ is well detected. A 20″ scale bar is shown for the H$_2$ image, which is displayed using a logarithmic intensity scale. The 10″ scale bars apply to the other three images. H$_2$ emission is found to be extended. Emission in the continuum subtracted H$_2$ image at the location of the Br$\gamma$ emission is most likely He I $3^3$P° $-$ $4^3$S at $\lambda = 2.114$ $\mu$m emission, which falls within the filter bandpass. b) NGC 3242 in H$_2$, $Ks$–band, Br$\gamma$, and continuum subtracted Br$\gamma$. H$_2$ is not detected. A 20″ scale bar is shown.

**Figure 5:** a) NGC 6210 shown in $\lambda = 2.121$ $\mu$m continuum (H$_2$ is not detected), $Ks$–band, Br$\gamma$, and continuum subtracted Br$\gamma$. A 10″ scale bar is shown. b) NGC 6302 in the UH $J$, H$_2$, Br$\gamma$, and continuum filters. A 20″ scale bar is shown. These images are shown on a logarithmic intensity scale.

**Figure 6:** a) NGC 6543 shown in the UH H$_2$, Br$\gamma$, continuum, and $I$–band filters. H$_2$ is not clearly detected. A 10″ scale bar is shown. b) NGC 6572 in the UH $K$, H$_2$, Br$\gamma$, and continuum filters. A 10″ scale bar is shown.

**Figure 7:** a) NGC 7009 in H$_2$, continuum subtracted H$_2$, $\lambda = 2.151$ $\mu$m continuum, and $K$–band. H$_2$ is not detected. These images are shown on a logarithmic intensity scale. A 10″ scale bar is shown. b) UH continuum and Br$\gamma$ images of NGC 7009. The bottom frames are shown on a logarithmic intensity scale. A 10″ scale bar is shown.

**Figure 8:** a) NGC 6826 (lower) in the UH $I$–band filter and shown at different contrast levels. UH $I$–band images of NGC 7026 (upper), also at different contrast levels. These images are displayed on a logarithmic intensity scale. A 10″ and 20″ scale bar is shown for each object respectively. b) NGC 7026 shown in H$_2$, continuum subtracted H$_2$, Br$\gamma$, and continuum subtracted Br$\gamma$. H$_2$ is not detected. The emission remaining in the continuum subtracted H$_2$ images is likely from He I. A 10″ scale bar is shown. The Br$\gamma$ images are shown on a logarithmic intensity scale.

**Figure 9:** a) NGC 7662 shown in $\lambda = 2.121$ $\mu$m continuum (H$_2$ is not detected), $K$–band, Br$\gamma$, and continuum subtracted Br$\gamma$. A 20″ scale bar is shown. All images are displayed on a logarithmic intensity scale. a) NGC 7662 in UH Br$\gamma$, continuum, and $I$–band. These images are shown on a logarithmic intensity scale. The $I$–band frames are at different contrast levels. A 20″ scale bar is shown.

**Figure 10:** a) NGC 7027 shown in H$_2$, continuum subtracted H$_2$, Br$\gamma$, and continuum subtracted Br$\gamma$. A 10″ scale bar is shown. b) BD+30°3639 (upper) and IRAS 21282+5050 in the UH $K$–band. The left images are shown on a linear intensity scale, while the right are on a logarithmic scale. A 5″ scale bar is shown for BD+30°3639 and a 5″ scale bar is shown for IRAS 21283+5050.

**Figure 11:** a) IC 289 (lower) shown in Br$\gamma$ and $K$–band. A 20″ scale bar is shown. IC 2149 (upper) in H$_2$ and continuum subtracted H$_2$. The remaining emission in the continuum subtracted image is from He I. A 10″ scale bar is shown for this object. b) IC 418 in the SO $K$, $H$, and UH $J$ bandpasses with Br$\gamma$. A 10″ scale bar is shown. The continuum contribution to the Br$\gamma$ image is moderate, and continuum subtraction (not shown) does not offer significantly different morphological information.

**Figure 12:** a) IC 4593 shown in $\lambda = 2.121$ $\mu$m continuum (H$_2$ is not detected), $K$–band, Br$\gamma$, and continuum subtracted Br$\gamma$. A 10″ scale bar is shown. b) J 900 in H$_2$, continuum subtracted H$_2$, continuum subtracted Br$\gamma$, and $\lambda = 2.151$ $\mu$m continuum. H$_2$ is clearly detected. It is expected that some of the emission found in the continuum subtracted H$_2$ image at the location of Br$\gamma$ is from He I in the bandpass. The 20″ scale bar applies to the top frames. The 10″ scale bar applies to the bottom frames.

**Figure 13:** a) M 57 in H$_2$, $\lambda = 2.151$ $\mu$m continuum, Br$\gamma$, and continuum subtracted Br$\gamma$. H$_2$ is well detected. A 40″ scale bar is shown. The H$_2$ image is on a logarithmic intensity scale. b) AFGL 618 shown in continuum subtracted H$_2$, continuum subtracted Br$\gamma$, and in the $J$ and $K$ bandpasses. A 5″ scale bar is shown. The bottom two frames are shown on a logarithmic intensity scale.

**Figure 14:** a) AFGL 2688 in H$_2$, continuum subtracted H$_2$, $\lambda = 2.151$ $\mu$m continuum, and in the UH $K$–band. The reduced 10″ scale bar applies to the $K$–band frame only. These data are discussed in detail elsewhere (Latter et al. 1993). b) M 1-92 shown in the UH $J$– and $K$–bands. The upper images are shown on a logarithmic intensity scale. The lower are on a linear scale. A 5″ scale bar is shown.

**Figure 15:** a) M 2–9 in the UH $H_2$, Br$\gamma$, continuum, and $K$–band filters. A $10''$ scale bar is shown. A detailed discussion can be found in Hora & Latter (1994). b) M 1–78 in the UH $K$–band. the intensity scale is logarithmic, with the frames shown at different contrast levels. A $10''$ scale bar is shown.

**Figure 16:** Contour plot of NGC 2346 in the SO $H_2$ filter. Contours are from $5.6 \times 10^{-16}$ to $8.3 \times 10^{-15}$ erg cm$^{-2}$ s$^{-1}$ arcsec$^{-2}$ with an interval of $5.6 \times 10^{-16}$ erg cm$^{-2}$ s$^{-1}$ arcsec$^{-2}$. The continuum is weak, and has not been subtracted from this image.

**Figure 17:** Contour plot of M 57 in the SO $H_2$ filter. Contours are from $1.9 \times 10^{-16}$ to $5.3 \times 10^{-15}$ erg cm$^{-2}$ s$^{-1}$ arcsec$^{-2}$ with an interval of $5.6 \times 10^{-16}$ erg cm$^{-2}$ s$^{-1}$ arcsec$^{-2}$. The continuum is weak, and has not been subtracted from this image.

**Figure 18:** a) Contour plot of NGC 7027 in the SO $H_2$ filter. Contours are from $2.5 \times 10^{-15}$ to $1.4 \times 10^{-13}$ erg cm$^{-2}$ s$^{-1}$ arcsec$^{-2}$ with an interval of $4.2 \times 10^{-15}$ erg cm$^{-2}$ s$^{-1}$ arcsec$^{-2}$. The bright inner ring is from He I $3^3P^\circ - 4^3S$ at $\lambda = 2.114$ $\mu$m emission, which falls within the filter bandpass. However, the more extended emission is $H_2$. b) As in a) but for the SO Br$\gamma$ filter. Contours are from $1.1 \times 10^{-14}$ to $1.0 \times 10^{-12}$ erg cm$^{-2}$ s$^{-1}$ arcsec$^{-2}$ with an interval of $2.8 \times 10^{-14}$ erg cm$^{-2}$ s$^{-1}$ arcsec$^{-2}$. These images have been continuum subtracted.

**Figure 19:** Contour plot of J 900 in the SO $H_2$ filter. Contours are from $1.4 \times 10^{-16}$ to $1.4 \times 10^{-15}$ erg cm$^{-2}$ s$^{-1}$ arcsec$^{-2}$ with an interval of $1.1 \times 10^{-16}$ erg cm$^{-2}$ s$^{-1}$ arcsec$^{-2}$. The continuum is compact and has not been subtracted from this image.

**Figure 20:** Contour plot of NGC 2440 in the SO $H_2$ filter. Contours are from $4.2 \times 10^{-16}$ to $2.6 \times 10^{-15}$ erg cm$^{-2}$ s$^{-1}$ arcsec$^{-2}$ with an interval of $2.8 \times 10^{-16}$ erg cm$^{-2}$ s$^{-1}$ arcsec$^{-2}$. The continuum is not extended with the $H_2$ emission, and has not been subtracted from this image.

**Figure 21:** a) Contour plot of AFGL 618 in the SO $H_2$ filter with continuum subtracted. Contours are from $2.8 \times 10^{-15}$ to $2.2 \times 10^{-13}$ erg cm$^{-2}$ s$^{-1}$ arcsec$^{-2}$ with an interval of $8.3 \times 10^{-15}$ erg cm$^{-2}$ s$^{-1}$ arcsec$^{-2}$. b) As in a) for the SO Br$\gamma$ filter. Contours are from $1.9 \times 10^{-15}$ to $2.2 \times 10^{-13}$ erg cm$^{-2}$ s$^{-1}$ arcsec$^{-2}$ with an interval of $8.3 \times 10^{-15}$ erg cm$^{-2}$ s$^{-1}$ arcsec$^{-2}$.